# Magnetization reversal processes in ErFe$_2$/YFe$_2$ exchange spring multilayer studied by x-ray magnetic circular dichroism

G. B. G. Stenning[1], A. R. Buckingham[1], G. J. Bowden[1], R.C.C. Ward[2], G. van der Laan[3], L. R. Shelford[3], F. Maccherozzi[3], S. S. Dhesi[3] and P. A. J. de Groot[1]

[1]*School of Physics and Astronomy, University of Southampton, SO17 1BJ, United Kingdom*
[2]*Clarendon Laboratory, Oxford University, OX1 3PU, United Kingdom*
[3]*Diamond Light Source, Chilton, Didcot OX11 0DE, United Kingdom*

**Abstract**

X-ray magnetic circular dichroism at the Er $M_{4,5}$ edge is used to study the switching behavior of the hard ErFe$_2$ layers in an epitaxial [ErFe$_2$(70Å)/YFe$_2$ (150Å)]×25 exchange-spring superlattice. Magnetic hysteresis loops for the Er magnetization, at temperatures $T < 200$ K, reveal a switching behavior with a single type of irreversible switch corresponding to vertical exchange spring states. Experiments at $T > 200$ K reveal a crossover to a regime with two types of switching processes. Computational modelling for this system gives a semi-quantitative agreement with the experiment and reveals that the observed high temperature switching behavior is due to a spin-flop like reorientation transition. In contrast to conventional spin-flop transitions in antiferromagnets, in this exchange spring system the increase in anisotropy energy of the hard magnetic layers is overcome by the decrease in Zeeman energy of the soft layers. Computational studies also reveal the presence of transitions between vertical exchange spring and spin-flop states with a first-order character as well as continuous transitions between these states.

## 1. Introduction

Exchange-spring magnetic multilayers, which consist of alternating hard- and soft-magnetic layers, are attracting a great deal of attention due to potential applications in data storage media [1-3], permanent magnets [4,5], and MEMS [6-8]. Such multilayers have been proposed as superior data storage media since they offer additional flexibility in optimising magnetic properties. For instance, they provide a degree of decoupling between the coercivity, which determines the required write field, and the energy barriers required to avoid thermally activated data loss [1]. In summary, these systems allow us to overcome the "superparamagnetic" limit. For such applications, it is important to develop a detailed understanding of the magnetization switching in exchange spring structures.

In recent years, it has been shown that multilayer films of Laves phase REFe$_2$ (RE = rare earth) grown by molecular beam epitaxy (MBE) are excellent model systems for the study of exchange spring phenomena [9]. In earlier papers [10,11] the magnetic properties of MBE-grown ErFe$_2$/YFe$_2$ superlattices were reported and discussed. In particular, it was shown that at low temperatures ($T \leq 200$ K), the easy axis is near an out of plane $\langle 111 \rangle$-type crystal axis. This direction is determined primarily by the cubic crystal field interaction at the Er$^{3+}$ site which favors the $\langle 111 \rangle$ axes. However in MBE-grown (110) ErFe$_2$ films, there is an additional strain term to the magnetic anisotropy which favors the [110] crystal growth axis [12]. Consequently, at low temperatures the easy axis lies out of plane, between either the [111]–[110] or [11$\bar{1}$]–[110] axes.

From a magnetic point of view, the magnetic exchange interactions can be categorized as follows. The strongest coupling is that of the Ferromagnetic (FM) Fe-Fe exchange (~600 K), which determines the high Curie temperatures of all the REFe$_2$ compounds. Next, is the antiferromagnetic (AFM) Er-Fe exchange (~100 K). Finally, the Er-Er exchange is small and is neglected in the simulations described below. At $T = 0$ K the effective magnetic moments on the Er and Fe sites can be set at 9 $\mu_B$ and 1.5 $\mu_B$, respectively [13]. This approach is a simplification; band structure calculations show that in addition to the 3$d$ moments on the iron, there are induced 5$d$ moments at the RE-sites [14]. However, since the 5$d$ moments are driven primarily by the Fe 3$d$ sublattice, it is a reasonable approximation to use a discrete two component ferrimagnetic model, provided we ascribe $\mu_{Er} = 9\,\mu_B$ and $\mu_{3d} + \mu_{5d} = 1.5\,\mu_B$ [13]. The magnetic anisotropy of the ErFe$_2$/YFe$_2$ multilayer films is generated primarily by the Er$^{3+}$ ions [15]. Here there are two competing interactions: (*i*) the normal cubic crystal field interaction, arising from the interaction between the charge distribution of the RE 4$f$ shell and the electric gradients within the Laves phase crystal and (*ii*) the growth-induced magneto-elastic interaction [12]. Both are electrostatic in origin. In contrast to the anisotropy associated with the ErFe$_2$ layers, the anisotropy of the YFe$_2$ layers is two orders of magnitude smaller. Thus, in zero field, the ErFe$_2$/YFe$_2$ multilayer adopts a simple AF state for the net layer magnetization with the magnetic moment pointing out of the plane of the film but not perpendicular .

When a magnetic field is applied along the [110] growth-axis, magnetic exchange springs can form in the soft YFe$_2$ layers. In this paper we shall refer to such a spin-state as a *vertical exchange spring*, given that the Er spins are constrained by their anisotropy to lie close to the film normal.

From bulk magnetometry and anomalous Hall effect measurements at low temperatures [10], it is found that there is one irreversible switch of the hard layers, from the [111] axis to an [1$\bar{1}\bar{1}$] out of plan axis on the reverse side of the film, thereby resulting in a simple $M$–$B_a$ loop. Bulk magnetometry at higher temperatures, in excess of 200 K, reveals additional irreversible switching at high fields, indicative of a type of reorientation transition. Computational work suggests that the reoriented state is similar to a spin-flop state in antiferromagnets and hence we will refer to this state as the *spin-flop state* [10]. Both the vertical exchange spring and the spin-flop state transition are likely to minimize their energies by exploiting <111> easy cubic crystal field directions.

X-ray magnetic circular dichroism (XMCD) measurements on DyFe$_2$/YFe$_2$ superlattices [16,17], have demonstrated the power of this technique for element specific magnetometry in the Laves phase materials. In the case of DyFe$_2$/YFe$_2$, which has an in-plane magnetic anisotropy, XMCD has revealed intricate magnetization reversal processes, in good agreement with computational studies [18].

Previous results from bulk magnetometry [10] on ErFe$_2$/YFe$_2$ superlattices showed only weak signatures of high field reorientation processes. By using XMCD the complex spin-reorientations and temperature-dependent magnetic reversal phenomena can be investigated in detail. In particular the Er XMCD enables the study of the magnetization ($M_{Er}$) behavior of the hard layer where the anisotropy resides, thus is likely to offer profound information about switching processes.

## 2. Experimental Procedure

The ErFe$_2$/YFe$_2$ superlattice samples were grown by MBE following the procedure detailed in Refs. [19]. A 100 Å Nb buffer layer and 20 Å Fe seed layer were deposited onto an epi-prepared $(11\bar{2}0)$ sapphire substrate of size 10 mm × 12 mm. The Laves phase film consisted of a superlattice of [ErFe$_2$ (70Å)/YFe$_2$ (150Å)]×25 grown with a (110) film plane and the major axes parallel to those of Nb. This was achieved by co-deposition of elemental fluxes at a substrate temperature of 600°C. To prevent the oxidation of the multilayer, the sample was capped with a 100 Å layer of Y.

XMCD measurements were performed on the soft x-ray undulator beamline I06 at Diamond Light Source. The beamline is equipped with a superconducting magnet in a high vacuum of $10^{-10}$ mbar. Field sweeps were conducted between ±6 T in the temperature range 150 K–250 K to observe the exchange-spring driven magnetization transitions. The magnetic field was applied parallel to the x-ray beam. In initial experiments the film normal was aligned 10° away from the x-ray beam direction. This geometry allowed for both x-ray fluorescence and electron drain current detection of the absorption, but yielded XMCD data which was identical for both methods. To simplify the analysis of the experimental data, the sample was reoriented with the film normal parallel to the magnetic field and x-ray beam. Due to the position of the Si photodetector in this geometry, fluorescence yield was impractical and only drain current collection was used for the experimental data presented in this paper.

The XMCD signal was obtained as the difference in absorption between left- and right- circularly polarised x-rays. This difference signal, the magnetic dichroism, can be used to deduce the magnetization of different elements in the sample, depending on the absorption edge used. The XMCD results were obtained using the Er $M_{4,5}$ ($3d \rightarrow 4f$ at 1405, 1440 eV) and Fe $L_{2,3}$ ($2p \rightarrow 3d$ at 710, 722 eV) transitions [20,21]. These allow for element specific hysteresis loops of the superlattice sample to be obtained. Experiments at the Y $M_{2,3}$ (300, 310 eV) absorption edge did not show any magnetic circular dichroism; which we attribute to the presence of the non-magnetic Y capping layer, dominating the signal and dwarfing any magnetic information from the YFe$_2$ layers.

Pronounced Fe XMCD was observed, but high temperature experiments of the Fe magnetization curves, $M_{Fe} - B_a$, revealed only weak features. This is consistent with the bulk magnetometry data on the ErFe$_2$/YFe$_2$ system.

## 3. Er XMCD Results

Results of the Er XMCD as a function of magnetic field ($B_a$) at 150 K are presented in Fig. 1(a), showing a relatively simple hysteresis loop with irreversible switching at 2.3 T. This data is consistent with the results of bulk magnetometry at low temperatures, showing that below the crossover temperature regime the magnetization undergoes a single switching event. The arrows in the figure refer to the direction of the field sweep. The gradual change of $M_{Er}$ near $B_a = 0$ can be explained by the small rotation of the magnetization from the easy axis to the film normal.

As the temperature is raised, a crossover between the single-switch of Fig. 1(a) to a double-switch behavior is observed in the magnetic hysteresis loops. This crossover changes drastically the Er magnetization curves as shown in Fig. 1(b) for $T = 220$ K.

Upon further increase of the temperature, the switching processes become better defined and the $M_{Er}$ hysteresis loops reveal two distinct irreversible switching events. For instance, at 250 K [Fig. 1(c)] the low field transition takes place at relatively low fields, 0.4 T, and the second transition at 3 T. The value of the high field transition is found to increase with increasing temperature.

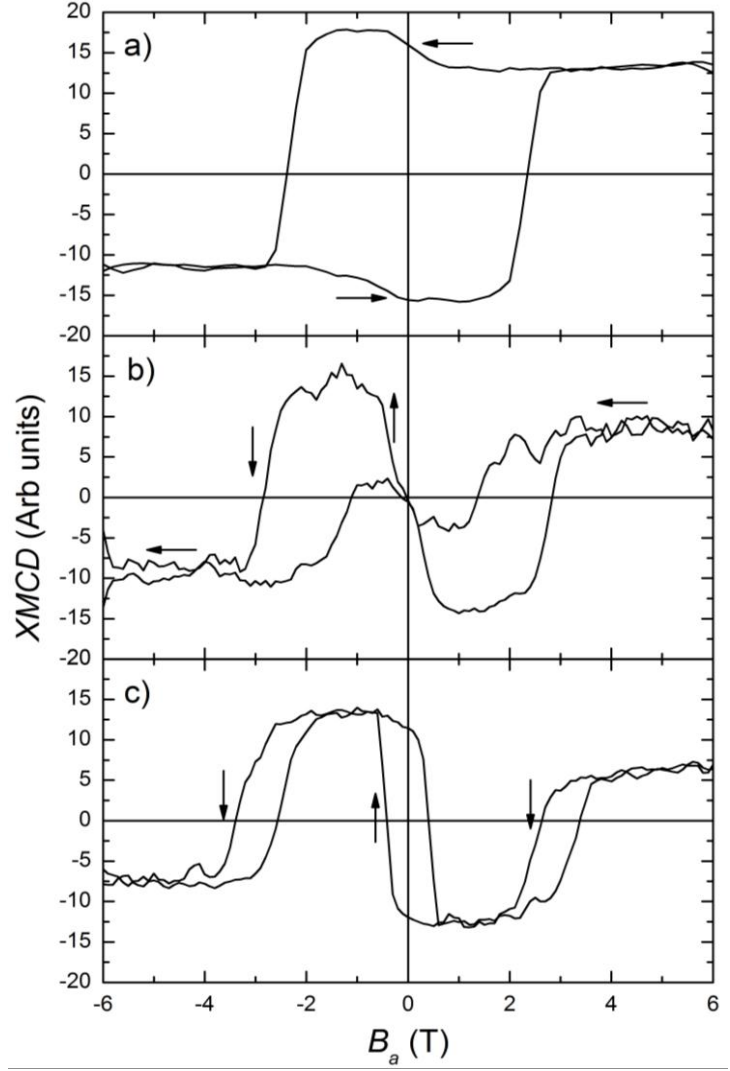

FIG. 1: $M_{Er}$ hysteresis loops obtained by XMCD using total electron yield at (a) 150 K, (b) 220 K, and (c) 250K.

## 4. Micromagnetic Simulations

To gain insight into the mechanisms which underlie the switching behavior, micromagnetic modelling was performed using the discrete 1-D model described in Ref. [13]. The parameters used in this model were established independently as a function of temperature as described in Refs. [22,23]. Parameters were not adjusted to fit the XMCD data. In this model stable magnetization arrangements are determined which correspond to energy minima. In a complex magnetic system, such as an exchange spring magnet, usually several stable arrangements are found for a given field and temperature. There are two methods by which switching can be determined in this model. In the first method the stability field range of a particular state is determined. On going beyond the stability range the magnetic system has to switch into another state which is stable at that field value. This process resembles in some way the Stoner-Wohlfarth model in that it determines the magnetic fields at which states become unstable as an indication of the switching fields, except that in the present method the magnetization is usually not uniform. Similar to the case of the Stoner-Wohlfarth model, the value found for the irreversible switching field, $B_{i,max}$, is the maximum possible value. In these computational studies a lower limit of the switching field can be found by assuming the system switches when a lower energy state becomes available, $B_{i,\min}$. The simulations revealed that, while in a simple ferromagnet $B_{i,\min}$ and $B_{i,\max}$ are very different (*e.g.*, $B_{i,\min} = 0$), in exchange spring multilayers the effect of the spring structure (which acts as a built-in domain wall) the two values are often comparable. As a working model we have taken values of the switching field to be the average of these two limiting cases. In practise we found that this provides values of the switching field in good agreement with experiment. For example, both $B_{i,\max}$ and $B_{i,\min}$ are shown in Fig. 2(a). In Fig. 2(b) and for the remainder of the simulation data we show the average values of the switching fields only.

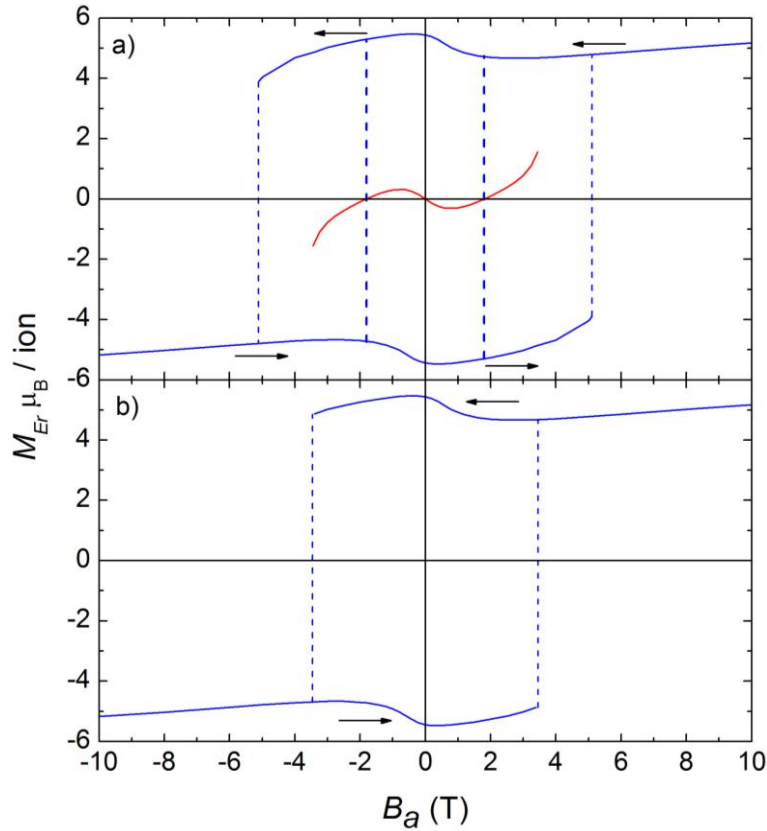

FIG. 2: (Color online) (a) Simulated $M_{Er}$-$B_a$ magnetic hysteresis loop at 150 K. Solid blue lines correspond to vertical (spin-flop) exchange spring states. The vertical outer dashed lines at 5.1 T (blue) correspond to $B_{i,\max}$. The inner dashed lines (blue) at 1.8 T are where the energy of both the two spring states is identical, $B_{i,\min}$. The solid red line at $B_a = 0$, $M_{Er} = 0$ gives the $M_{Er}$ stable spin-flop states. (b) Simulated $M_{Er}$-$B_a$ loop at 150 K using the average switching field $B_c = 3.58$ T.

The simulations revealed two types of states for applied fields normal to the sample plane. The first is the vertical exchange spring state as described above. It is found that for this state all the moments (both Er and Fe ones) are confined to the $(\bar{1}10)$ plane, *i.e.*, possessing no $[\bar{1}10]$ components. In the second type of state, the reorientation state, components of the moments out of the $(\bar{1}10)$ plane develop and the Er moments are directed away from the film normal. The direction of the average Er moment is illustrated in Fig. 3.

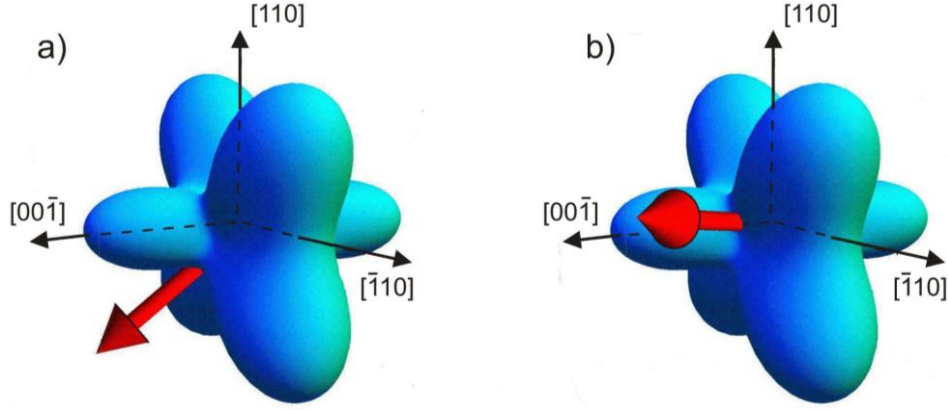

FIG. 3: (Color online) Average Er moment obtained from the simulations with respect to the single ion anistropy energy for (a) the vertical exchange spring and (b) the spin-flop state.

The calculated Er magnetic loop at 150 K (Fig. 2) show both the vertical exchange spring (solid blue) and spin-flop states (solid red at $B_a = 0$, $M_{Er} = 0$). The reorientation state is higher in energy than its vertical exchange spring counterpart. Such a state is therefore metastable between -3.4T and +3.4T from the calculation.

The $B_{i,max}$ prediction for the switching field is 4.78 T. This is indicated by the blue dotted vertical line in Fig. 2(a). The energy crossover between the vertical exchange state and the corresponding reversed state occurs at $B_{i,min}$ = 3.36 T. Using the average switching field we obtain the magnetization loop shown in Fig. 2(b). Hence the low temperature $M_{Er} - B_a$ loop is characterised by a single magnetic switch at 3.58 T. This switch corresponds to a transition between states with out of plane <111> alignment of the Er moments, from one side of the plane to the other. As these multilayers are dominated by the magnetization of the soft layers, the magnetization switches into a direction roughly antiparallel to the applied field. The Er XMCD loop obtained at 150 K presented in Fig. 1(a) shows striking similarity to the micromagnetic data in Fig. 2(b). Note that both show the increase in $M$ as the applied magnetic field passes through zero.

In going from 150 K to 200 K the overall shape of the simulated $M_{Er} - B_a$ loop changes dramatically. At the higher temperature, there are now two distinct magnetic switching fields. Once again we have used the simple average switching fields to produce the magnetization curves shown in Figs. 4(a) and 4(b). As the temperature is increased, micromagnetic simulation predicts that there should be a crossover from single-switch to a double-switch magnetic loop, as shown in Fig. 4(b). In the XMCD experiments this occurs at 220 K, however, the best agreement is found for simulation data at slightly lower temperatures. This difference can probably be attributed to the inaccuracy in the values used for Er anisotropy parameters.

It must be noted that, while at low and high temperatures the agreement between experimental and simulation data is good, in the crossover regime there is only qualitative agreement in the shapes of the magnetization curves. In this regime the simulated data shows relatively little hysteresis while this in not the case for the Er XMCD curves. This behavior may be due to the anisotropy associated with the $YFe_2$ layers or the dipolar interaction, which we have neglected and may cause some disagreement between theory and experiment, in particular at these temperatures where the energies of the different states are relatively balanced.

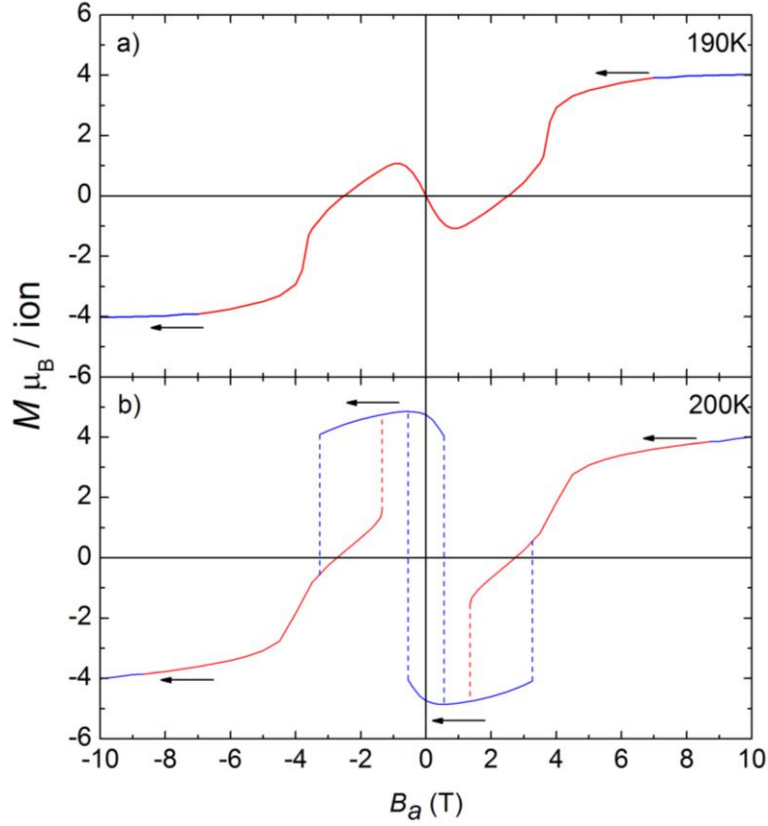

FIG. 4: (Color online) Simulated Er magnetization curves at (a) 190 K and (b) 200 K.

## 5. Instabilities

Spin configurations associated with particular states become unstable at certain temperatures and fields. In the computational studies such singular points are established by examining the double-derivative matrix $\mathbf{E}''_{total}$ of the total energy $\mathbf{E}_{total}$, as detailed in Ref. [24]. If all eigenvalues of $\mathbf{E}''_{total}$ are positive, the spin configuration defined by the set of angles, $\theta_i$ and $\phi_i$, of the magnetization in the simulation cells represents a stable spin configuration. Here, $\theta$ is defined as the inclination angle with respect to the [110] growth axis and $\phi$ is azimuthal angle between the projection of the magnetic moments in the film plane and the in-plane [001] axis. On the other hand, if any of the eigenvalues is negative, the spin configuration is unstable representing energy maxima rather than minima. Hence this offers a mathematical method to identify the onset of instabilities. Such data has been used to investigate the crossover from pure, vertical exchange spring behavior to more complex processes. The field dependences of the average inclination angle for the Er moments, $\langle\theta_{Er}\rangle$, at 190 K and 220 K are shown in Fig. 5. The data for 190 K reveals a continuous curve: the vertical exchange spring state is stable for all fields studied. However, if the temperature is raised above the crossover, a field range appears in which this state is not stable. Hence the magnetic configuration rearranges in another stable magnetization state which is available to the system; in this case the spin-flop state.

To provide further insight into the nature of the instability, we examined the spin-flop instability for 200 K. In Fig. 6, the average angles $\langle\theta_{Er}\rangle$ and $\langle\varphi_{Er}\rangle$ of the Er moments are shown as a function of magnetic field at 200 K. For decreasing fields the system undergoes a discontinuous (first-order) transition at 1.3 T. The parameter $\langle\varphi_{Er}\rangle$ acts as an order parameter distinguishing the two states:

$\langle \varphi_{Er} \rangle = 0$ for the vertical spring state and $\langle \varphi_{Er} \rangle \neq 0$ for the spin-flop state. In addition to the transition at 1.3 T a continuous transition between the two different states occurs at 8.7 T.

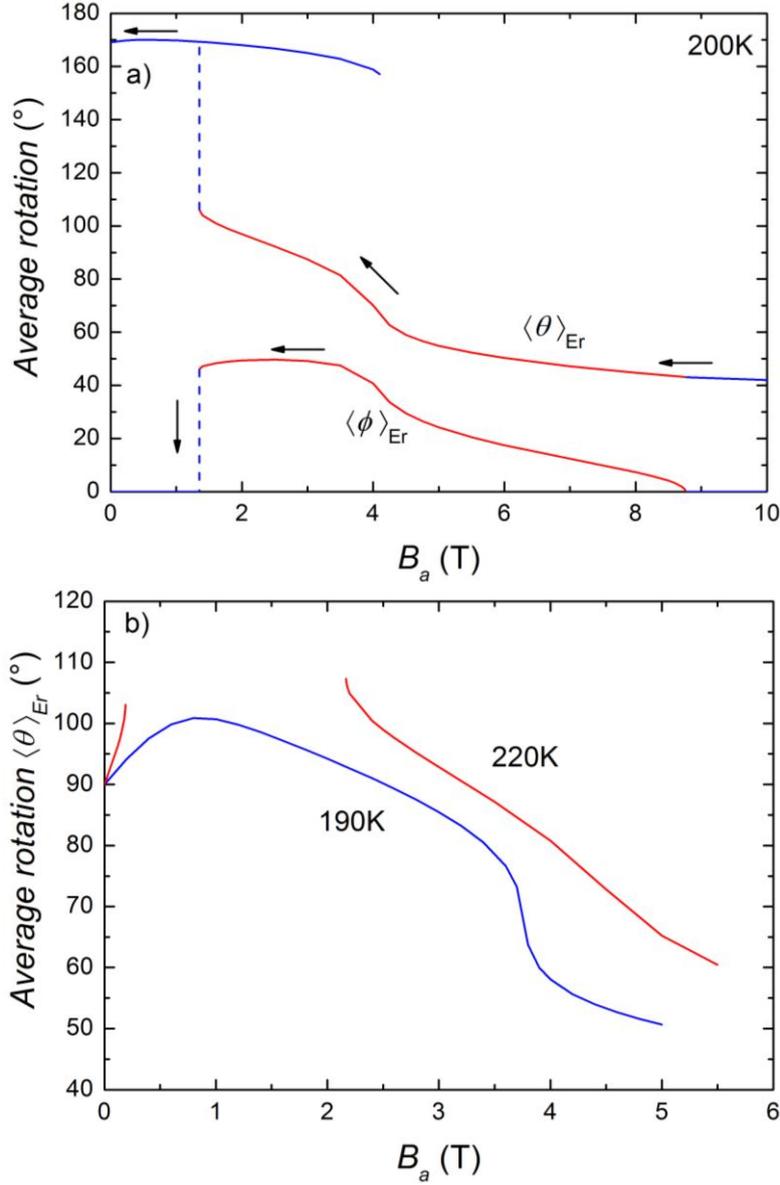

FIG. 5: (Color online) (a) Average $\langle \theta \rangle_{Er}$ and $\langle \varphi \rangle_{Er}$ corresponding to stable states at $T = 200$ K. (b) Average $\langle \theta \rangle_{Er}$ at 190 K and 220 K.

Investigation of the relationship between the direction of the average Er moment and its associated anisotropy field, revealed that at 200 K, for an applied field decreased from high fields to 3 T, the average Er moment is located in a small energy minimum, associated with an in-plane <111>-type axis. But as the field is reduced to 1.3 T, the Er moment rotates steadily downwards ($\langle \theta \rangle_{Er}$ increasing, $\langle \varphi \rangle_{Er}$ decreasing), eventually overriding the small energy saddle point between the in-plane and out of plane <111> type axes. When this occurs the spin configuration will collapse into a vertical spring state, as observed in the XMCD data. At 220 K, the energy barrier in question is small. We find that the anisotropy barrier is $\Delta E = 0.051$ K per Er ion. However at 190 K, the energy barrier has increased to $\Delta E = 0.182$ K per Er ion. This threefold increase is sufficient to trap Er moments in the energy minimum associated vertical exchange spring state in magnetic hysteresis loops. Clearly, the topological shape of the $Er^{3+}$ single ion anisotropy curve plays a major role in the pinning/depinning of the exchange spring and spin-flop states.

## 6. Conclusions

XMCD measurements at the Er $M_{4,5}$ edge using electron drain current detection have been performed yielding valuable information about the switching processes in $ErFe_2/YFe_2$ exchange-spring superlattices. It was found that magnetization switching transitions at high temperatures, which are insignificant in bulk magnetometry data due to the nearly-complete cancellation of the magnetization of hard and soft layers [10], are pronounced in the magnetization curves of the Er ions. The experiments confirms the crossover from a relatively simple single switching behaviour for $T < 200$ K to double switching for $T > 220$ K, inferred from bulk magnetometry.

Micromagnetic modelling, using a 1-D model for finding minimum energy states, gives a semi qualtitative agreeement with the XMCD data. The origin of the double switching has been traced to instabilities in *both* the spin-flop configuration and vertical spring configuration, at temperatures ~200 K. These, in turn can be associated with the topological shape of the $Er^{3+}$ single-ion anisotropy. Depinning of exchange spring states is determined primarily by low-energy saddle points which allow the Er magnetic moments to change direction without riding up over the much larger anisotropy maxima.

The model shows that the magnetic states associated with the switching event can be classified in two types. At low temperatures, in standard magnetic hysteresis loops, only vertical exchange spring states are traversed; all the magnetic moments are confined to the $(\bar{1}10)$ plane. However for temperatures above a crossover temperature regime around 200 K, other magnetization states play a role. These states have components of the magnetic moments pointing out of the $(\bar{1}10)$ plane and have a hard-layer magnetization which prefers directions close to the film plane. These states has a strong similarity to spin-flop states in classical antiferromagnets, in which for both sublattices an increase in anisotropy energy is overcome by a decrease in Zeeman energy. In contrast, in the reorientation transition in $ErFe_2/YFe_2$ superlattices the increase in anisotropy is associated with the hard layers and the decrease in Zeeman energy is due to the exchange spring structure in the soft layer consequently, exchange spring multilayers can undergo an exchange-spring driven spin-flop transition. In summary, hard-layer specific magnetometry enabled by XMCD, combined with micromagnetic modelling offers valuable opportunities to investigate magnetization states and switching processes in complex multilayer systems.